# Single step precursor free synthesis and characterisation of stable Au nanochains by laser ablation

By *Rajesh Kumar Neogy*[1], *Rajib Nath*[1], *Gautam Basu*[2] *and A. K. Raychaudhuri*[1]

[1]Department of Material Sciences
S. N. Bose National Centre for Basic Sciences,
Block-JD, Sector-III, Salt Lake, Kolkata-700 098(India)
[*] Email: neogy@bose.res.in
rajibn@bose.res.in
arup@bose.res.in

[2]Department of Biophysics
Bose Institute, Centenary Campus
P 1/12, C. I. T. Road, Scheme - VIIM
Kolkata –700054(India)

In this paper we report a simple one step and one-pot synthesis of stable assembly of Au nanoparticles (diameter 8-10nm) into chains in an Ethylene Glycol medium, using only a solid metallic Au target and a pulsed excimer laser (λ=248nm). The process reported does not use any precursor, reducing agent or surfactant and thus can be described as chemistry free synthesis route. The Au nanoparticle-ethylene glycol nanochains (with unbroken lengths often more than few microns) formed in liquid medium are mechanically as well as thermally stable and can be transferred unchanged into a solid substrate which can span a large surface area. The nanochains show a broad optical absorption covering almost the complete visible spectrum. A hybrid consisting of Au nanochains and separated nanoparticles can be formed by the same method using a proper choice of the laser fluence and Ethylene Glycol /DI water concentration. The Au nanochain - Ethylene Glycol hybrid material formed by the above method shows enhanced low frequency dielectric constant (one order more than the Ethylene Glycol) and enhanced electrical conductivity even with low Au fill fraction can show enhancement up to two orders. Based on 2D-NMR experiments we suggest that the quasi-1D chain like structure forms due to the formation of dimer and trimers of Ethylene Glycol molecules that attach to the Au nanoparticles formed by the ablation process and facilitate the chain formation.

## 1. Introduction

Assembly of nanoparticles in different forms has attracted a great deal of attention because of their unusual optical, electronic and magnetic properties that are strongly dependent on the size and shape of the particles. Organization of nanoparticles into nanowires with a networked structure is an important morphology for assembling nanoparticles. In this context multi-functional Au nanoparticles assembled in different types of morphology is being pursued by different groups [1]. This is a challenge for materials scientists investigating functional materials to exploit the multi-functional properties of Au nanoparticles in assemblies. Gold is one of the metals with the most positive electron affinity and thus disperses easily. Alcohols (here MEG) were found to be the best solvent for this system. Therefore, these combinations give the most stable nanometre-sized colloidal dispersion. The motivation for using Au nanoparticles ranges from device application of nanoparticles to even biomedical applications (due to its biocompatibility) that include imaging as well cancer therapy. In recent years hybrid materials containing water soluble polymers (like Chitosan) attached to nanoparticles are being used for biomedical applications [2]. Particularly Au nanoparticles-Polyethylene Glycol (PEG) hybrid nanocomposites are finding *in vivo* applications in radiation therapy [3]. Au nanoparticles coated with polymers have been attached to Mono Ethylene Glycol (MEG) through chemical means [4]. Au and Pt nanoparticles within dendrimeric polymeric networks have been synthesized as novel hybrid materials [5, 6]. Important functionality in Au-PEG system that has not been explored is the enhancement of dielectric constants and electrical conductivity with rather dilute Au addition while retaining the fluidity of the Ethylene Glycol.

Formation of nanoparticle-chains requires a suitable method to organize them into a quasi-one dimensional patterns. Most of the reported methods use chemistry based routes that include precursors, reducing agent and a capping agent. There are two predominant path ways. In the first pathway one starts with preformed nanoparticles and then assembles or self –organize them into the desired structure. The other pathway, often referred to as one-pot synthesis route, the nanoparticles are created as well as assembled in a chain of sequential chemical steps that takes place in the same reactor.

Chen et al. [7] fabricated both networked Au nanostructured particles and twisted Au nanorods suspended in water via a laser ablation technique using Nd: YAG laser (532 nm radiation) and to stabilize sodium dodecyl sulfate (SDS) surfactant was added to the nanoparticles colloids. Mafune´ et al. [8] also reported a two step method for formation of a chain network of Au nanoparticles. Reports using laser [9] or intense X-ray source from a synchrotron [10] to make Au nanoparticles and chains have been made using precursor of Au ($HAuCl_4$). Though there are few reports on the synthesis of Au nanochains via chemical route but there are very few works reported so far on the electrical characterization of such nanocomposites and the dielectric enhancement of the nanofluid.

In this paper we report a rather simple one step and one-pot synthesis of stable assembly of Au nanoparticles into chains in MEG medium, using only a solid metallic Au target (no precursors or reducing agent) and a pulsed excimer laser. It is noted that the method, presented here to make stable Au nanoparticle-chain, has the novelty that it is a single step, uses no surfactant or precursor and no reducing agent. The fact that the process uses no chemicals other than the Au target and biocompatible Ethylene Glycol, the Au nanochains so formed are immediately ready to use for any biological applications without any further purifications. We have also studied the dielectric properties of the Au-PEG systems.

## 2. Experimental Method

*2.1. Au nanochains synthesis conditions*

Surfactant-free Au nanonetworks were prepared by laser ablation of a Au metal plate (>99.99%) which was placed on the bottom of a quality quartz cuvette (HELLMA GmbH, Germany) filled with 5mL of commercially available Mono Ethylene Glycol (GR grade from Merc. Chemicals Co.). After proper focussing the laser ablation was performed by using the fundamental (248nm) of KrF gas (Coherent laser, Germany) pulsed laser operating at 5 Hz and energy was 200mJ and number of laser shots given was about 5000. We have made samples by just varying the concentration of MEG in de-ionized (DI) water in the base liquid but the samples were made by the  keeping the laser fluence as well as the number of shots unchanged . The concentration of the base liquid varied from pure DI water (MEG 0%) to no water (100% MEG) and the corresponding samples with these base liquids are listed in table 1.

Table 1

| Sample No. | Conc. of MEG (%) |
|---|---|
| 1 | 0 |
| 2 | 25 |
| 3 | 50 |
| 4 | 75 |
| 5 | 100 |

*2.2. Characterization and measurements*

For morphological characterisation SEM, TEM, EDX images were taken using (FE-SEM, FEI HELIOS 600) using 30kV electrons and 200kV JEOL JEM 200CX respectively.

Optical characterisation of the colloids were done by, UV-vis spectra (400-700nm) of the samples were recorded on a Shimadzu UV-2408 spectrometer.

$^{1}$H-NMR experiments were performed in $H_2O/D_2O$ with TSP as the internal standard on a Bruker DRX 500 spectrometer at 25°C for finding the chemical structure involved in the nanochains.

The dielectric measurement of these colloidal samples was performed by using impedance spectroscopy (IS) in a sample holder type coaxial capacitor coupled in an impedance analyser (HIOKI 3535 LCR HITESTER impedance analyser).

**3. Results**

*3.1 Structure of the nanoparticles and the chain morphology.*

To investigate the real state of the solution the TEM images were obtained from the solution (with different MEG concentrations) without any purification or centrifugation. From the TEM images of sample 5 (figure 1) it can be seen that a network of Au nanoparticle forms and it extends over a length scale that is much larger than the size of the nanoparticles. The chains are continuous over a length scale of few hundred nanometres (typically 50-100nm), before they branch off. This recurring pattern of continuous length and branching off continues for length scales exceeding few microns. In figure 2 we show the HR-TEM image of a typical part of the Au chain network. The average diameter of the nanoparticles is approximately 8-10nm. It can be seen that the network contains Au nanoparticles that appear to be co-joined. The high resolution image in figure 2(b) shows the lattice spacing of 2.5 Å which corresponds to the lattice spacing of gold.

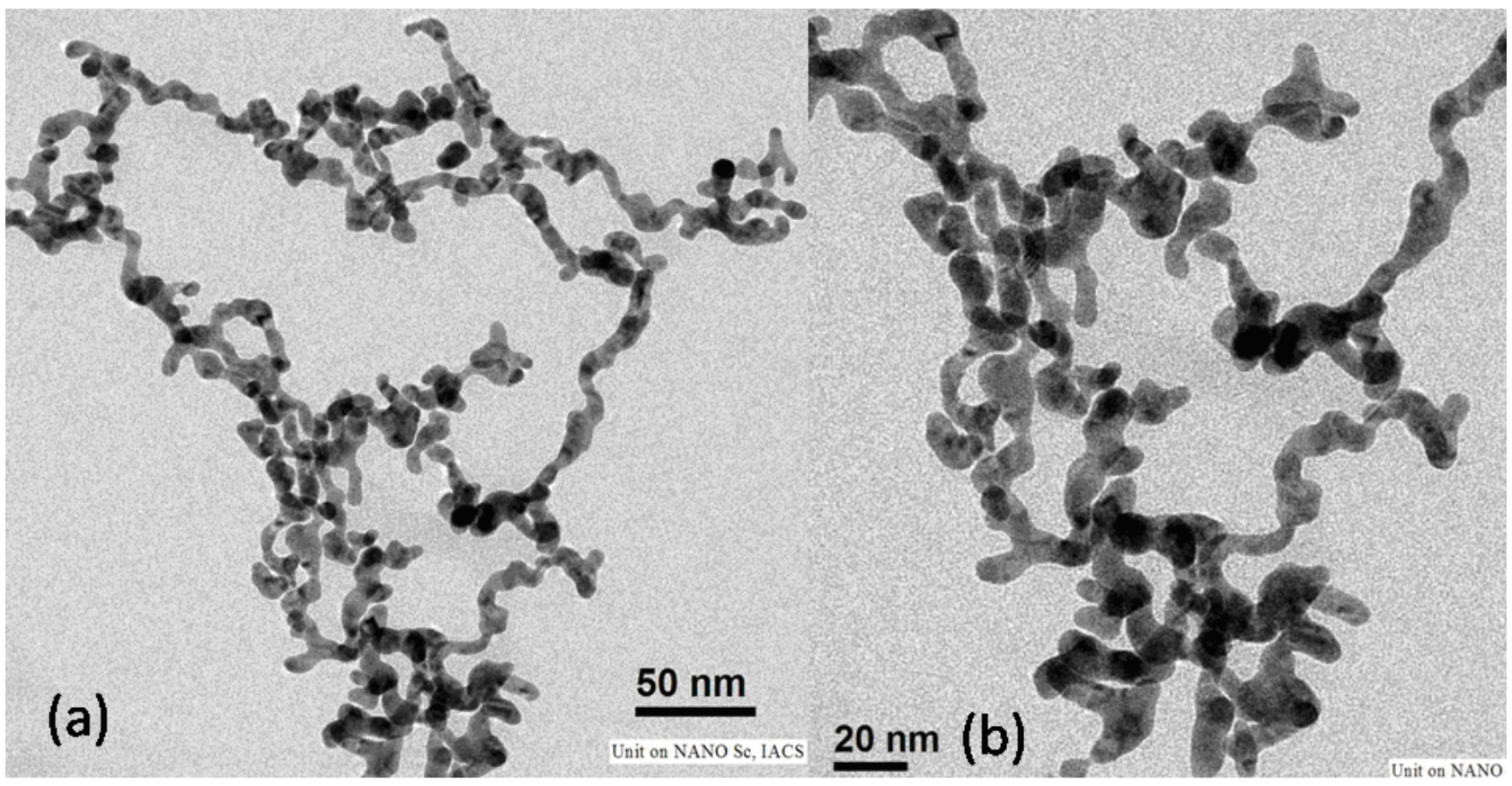

Figure 1 TEM image (sample 5) of the Au nanochain network at two length scales .

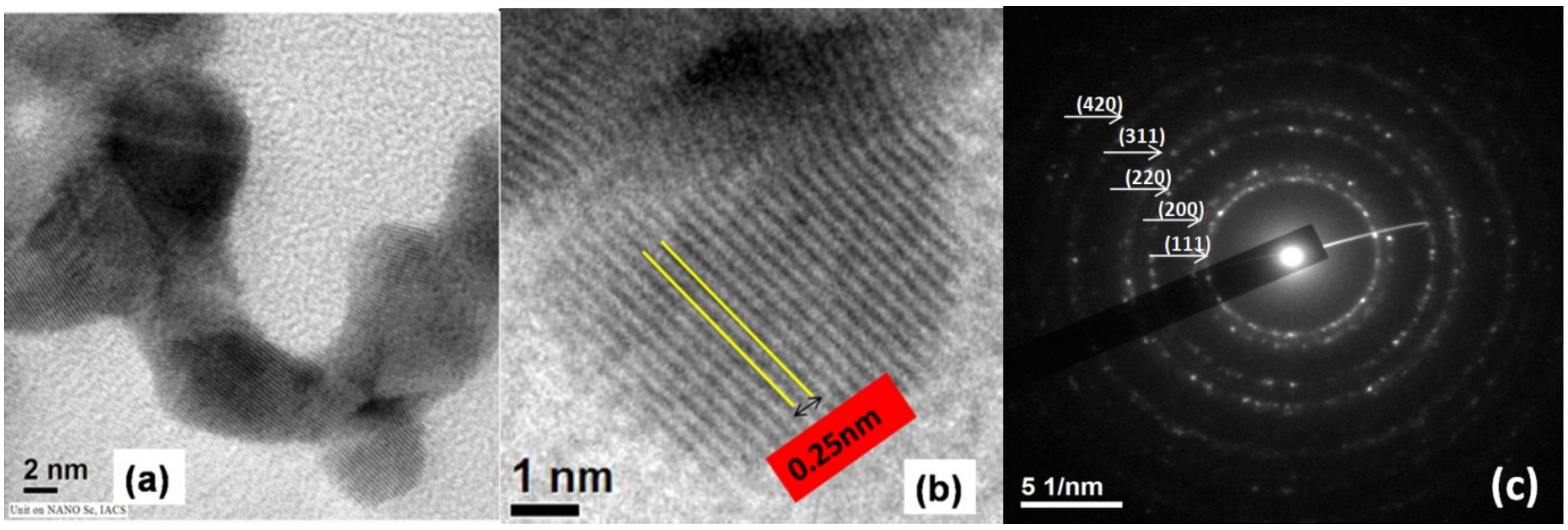

Figure 2 (a) HRTEM image of the Au nanoparticles in the chain. Individual nanoparticles cojoined to form the chain. (b) The lattice image of a single particle is shown.(c) SAED image.

The electron diffraction from typical particles in the chain is shown in figure 2(c), showing the scattering points corresponding to (111), (200), (220), (311), and (420) of Au crystalline (cubic) facets, confirming that Au nanochain network includes many small crystalline particles that have random and independent orientations. From the diffraction pattern we could find the lattice constant of the Au (bulk) to be 4.06 Å. The Energy Diffractive X-Ray Analysis (EDX) obtained from the SEM pattern show presence of only Gold. The analysis of the TEM data shows that the chain formation is not a random assembly or coagulation. In the event of random assembly this would lead to formation of random patches.

These network of nanochains spans over a large surface area when transferred to a solid surface. We were able to transfer the network on Si (100) wafer (with native oxide, typically 10nm thick) by drop coating. SEM Image in figure 3 shows the typical physical spread of the network of Au nanoparticle chain (sample 5) that spans over lengths more than 5μm or more and it can be transferred on to substrates like Si, Glass etc., so that it can be used to make devices from them using electrodes fabricated by lithography. As stated, the length scale of the organized structure is thus more than 3 orders of magnitude larger than the individual nanoparticles. A close examination of the nanostructure showed in the magnified scan shows that the region, where the network is formed, is rather dense. We have carried out tests (described below) that the nanoparticles that form by laser ablation of the Au target are mostly taken up by the network consisting of MEG. The process is thus an efficient process in terms of formation of the nanostructure.

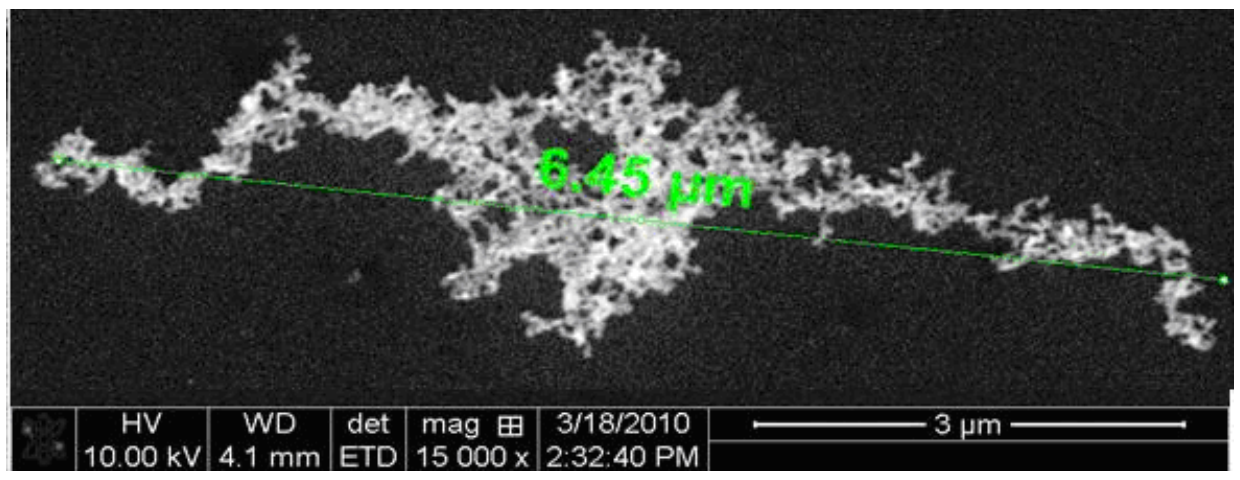

Figure 3 SEM images of the Au network (sample 5) on Si substrate

This network structure is rigid enough to survive (retains its structure) ultrasonic vibration (120W) in water after high-speed centrifugation at 5000rpm for about 15min. The rigidity of the network is thus preserved on mechanical handling and thus allows it to be transferred to a substrate for further studies and use. Its thermal stability has also been tested by heating the solution upto $70^0$C. No aging effect observed upto three months in normal conditions.

*3.2 Optical properties*

It is well known that, UV-Visible spectra can be correlated with the microstructure of the nanochains and helps us to establish the necessity of the MEG in formation of the nanochains as explained below. Figure 4(a) shows the optical images of the samples marked (1) to (5). The sample (6) represents the pure MEG (which is shown for comparison only).

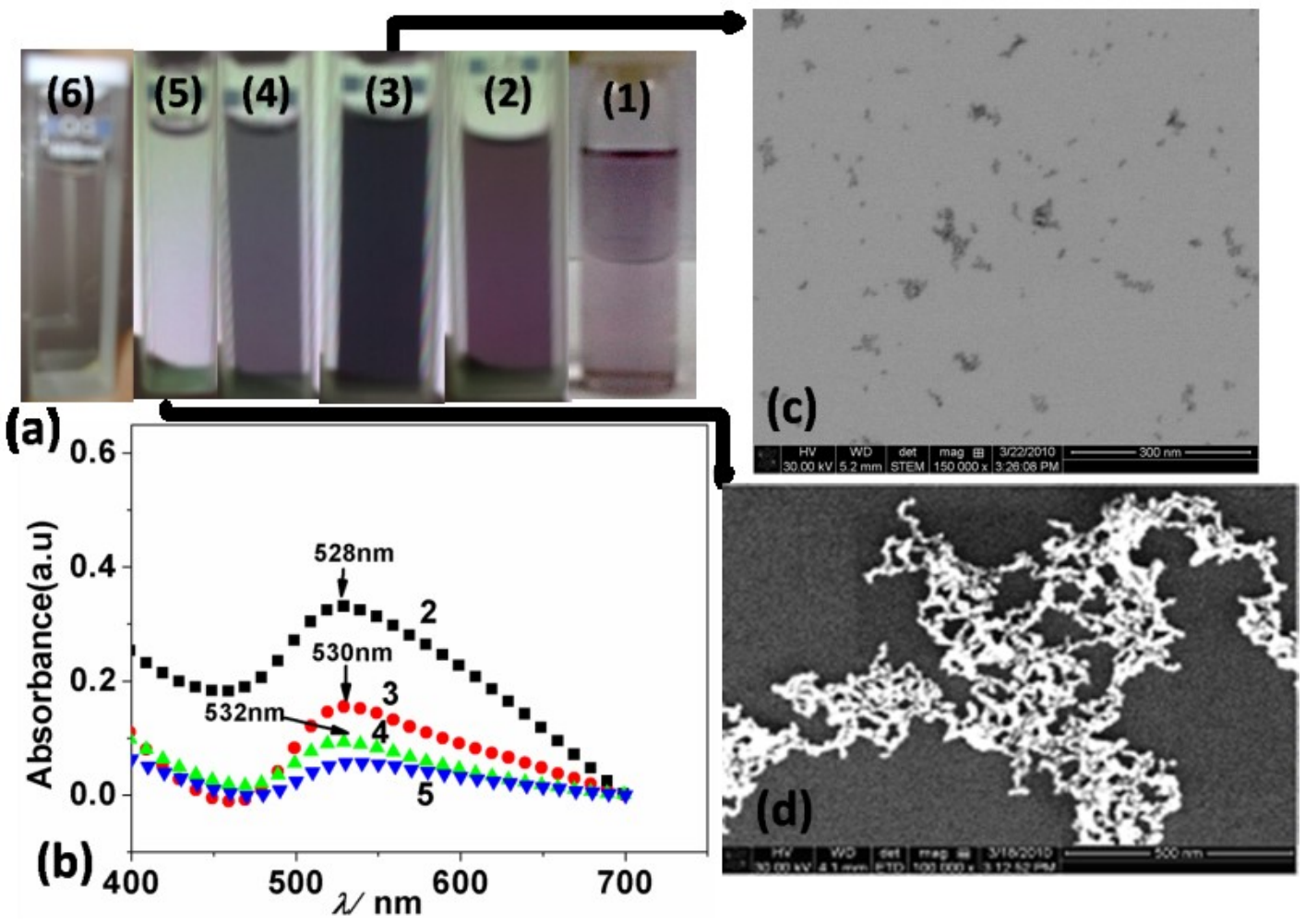

Figure (4) (a) Visual appearance of the samples, (b) Corresponding UV-Vis spectra, (c) & (d) are the SEM images of sample 3 & 4 respectively.

Figure 4(b) shows the variation of the UV-vis spectra for samples containing different proportion of MEG to DI water. As the proportion of MEG is gradually increased in the solution, there is a corresponding broadening of the absorption spectra to the long wavelength region. The change in the UV-Visible spectra on the Au chain formation, in particular red-shift of the surface plasmon band and its broadening on formation of chains, will be discussed later on. Interestingly the coexistence of two species can be seen in the UV-Visible absorption spectra that shows the 520 nm line (from nanoparticles) superimposed on a broad peak due to formation of chains. When the base liquid contains the MEG only (sample 5), UV-Vis spectra shows a board absorption extending from 500 to 700 nm and beyond [11] .The corresponding SEM data (figure. 4(d)), show that the chain formation is complete and total. In this case no plasmon peak is seen at 525nm arising from a single nanoparticle. The gradual increase of the network structure with MEG concentration indicates that it plays an essential role in the chain formation. There is a transformation from the isolated nanoparticles in pure water to chain formation and a network structure as the percentage of MEG is increased. This observation also allows us to tailor the absorption of the medium just by varying the relative concentration of the DI water to MEG in the base liquid.

Figure 5(a) shows the UV-Vis spectra for the sample 1 which has absorption spectra which peaks at 525nm. This corresponds to the surface plasmon resonance peak for isolated Au nanoparticles in the solution. The TEM picture of the sample (not shown here) also conforms the presence of isolated particles (some of which are agglomerated). In absence of MEG the sample thus contains only nanoparticles and no formation of chain has been detected.

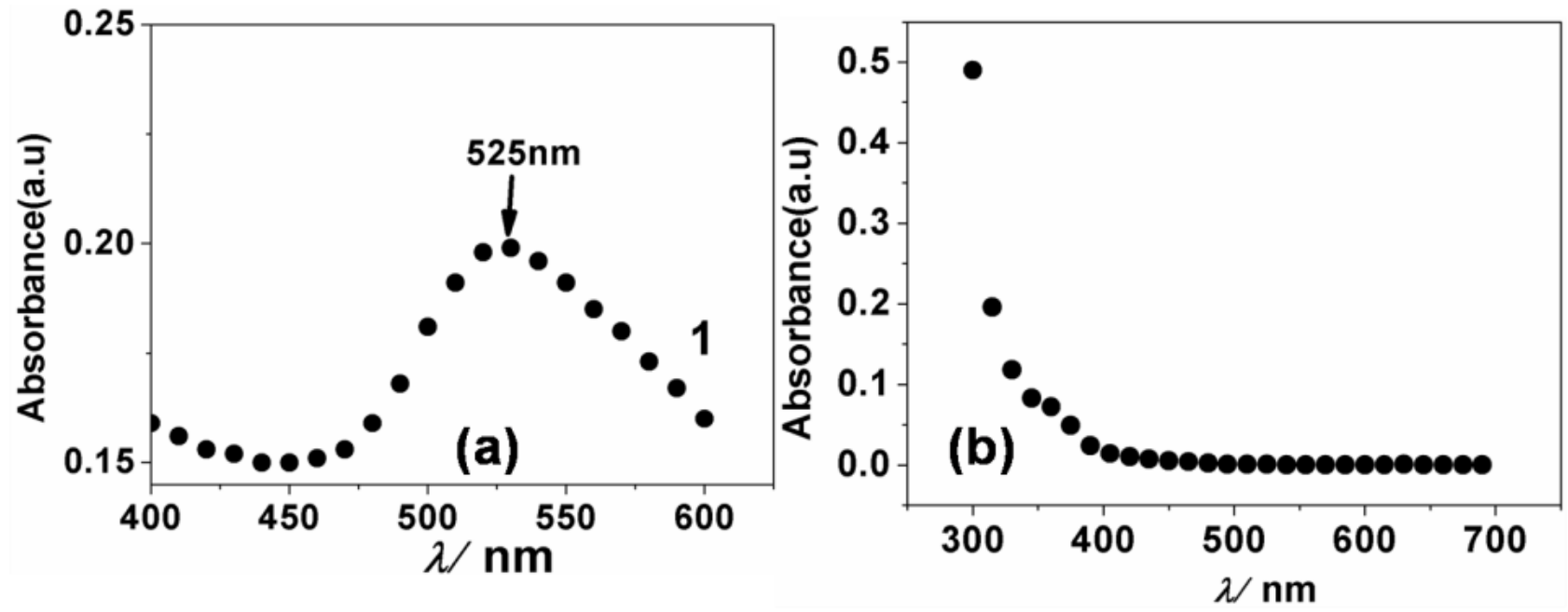

Figure (5) (a) UV-Vis spectra of sample (1) which has only DI water as base liquid. (b) UV-Vis spectra of filtered sample 5 after removal of Au chains by filtration.

One question that is important is whether the chains that are imaged on solid surfaces (Si substrates or TEM grid) form during solvent evaporation [12] or the chains exist in the base liquid as a product of the laser ablation. The UV-Visible absorption that develops on chain formation goes to show that these chains exist in the base liquid as a product of laser ablation and do not form due to solvent evaporation. As a further test that Au nanoparticles do exist as nanochains in the base liquid, we filtered the nanochains contained in the base liquid of sample 5 by a filter paper (WHATMAN) with 200nm pore size. Since most of the chains have dimensions in excess of 200nm, it is expected that the filter paper will arrest all the Au containing chains. The filtered base fluid without the Au nanochains network shows the absorption spectra (figure 5 (b)) which is the same as that of the parent MEG. The above test shows that the chain networks of size in excess of 200nm exist in the base fluid of MEG as a result of the ablation process.

The main change in the optical spectra on the chain formation is the change from a single particle type (sharp single frequency absorption band) to broad absorption spectra that has been significantly red shifted. Such a red-shift and broadening during chain formation has been investigated before by forming nanoparticles within pores of alumina templates using thermal decomposition of clusters of Au-complex [13]. One can use established theoretical descriptions of Mie scattering from similar aggregates. Such aggregates can show additional Plasmon resonance absorption at long wavelength in the optical absorption spectrum relative to the absorption from isolated nanoparticles dispersed in solutions [14]. This new long wavelength band is associated with the longitudinal mode of the electronic plasma oscillation along the long axis of the gold nanoparticles chains. Using extended Mie Scattering theory [15] red shift and broadening of the absorption can be explained and this is a direct manifestation of the chain formation as seen in the absorption spectrum.

*3.3 Dielectric property and resistance calculation of gold nanochains*

One of the new functionality that synthesized material shows is in their modified electrical property which shows much enhanced dielectric constant as well as much enhanced electrical conductivity. Impedance spectroscopy [16] (IS) which is a well established method to investigate dielectric/electrical properties of materials particularly multiphase systems. [17] The base fluid containing the MEG and the chain of Au nanoparticles (which will refer to as nano-composite) is a biphasic system containing heterogeneous mixture of a continuous phase (MEG) and a dispersed phase (Au nanochain network). The IS done as a function of frequency on the nano-composite sample allows us to separate out the contribution of the MEG matrix from that of the Au network. A simple model based on lumped circuit of capacitors and resistors (described below) was used for the analysis. The data for complex impedance Z were taken from 100Hz to 120MHz. At lower frequency the data is affected by electrode polarization and is thus not considered for analysis.

We have constructed the Cole-Cole plot (see figure 6) for the sample 5 and the corresponding base liquid (MEG). Both the plots are semi circular arcs in complex Z planes indicating a simple Debye type relaxation process. The frequency dependence of the impedance is arising from the time constant determined by the RC network only and not from the frequency dependences of the dielectric constant or the resistance which can be seen to be nearly frequency independent. From the plot itself, without any analysis it can be seen that the impedance of the composite is much lower than the MEG base fluid pointing towards a much enhanced conductance of the MEG-Au nanochains composite.

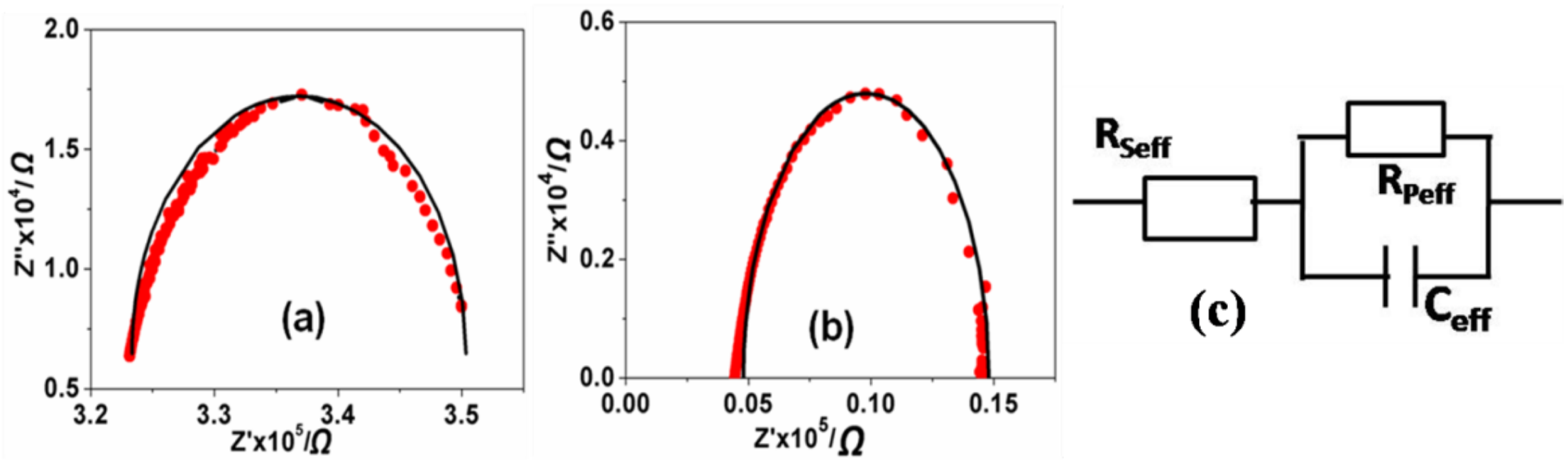

Figure (6) The Cole-Cole Plot for the MEG (a) and the MEG with the Au nanochains (b). The solid line is the fit to the simple model as described in text. (c) RC Circuits used for modelling the Cole Cole plot of the samples.

The real ($Z'$) and imaginary ($Z''$) part of the sample impedance can be expressed in eq. (1) from the above RC model:

$$Z' = R_{seff} + \frac{R_{peff}}{1+\omega^2 C_{eff}^2 R_{peff}^2} \,; Z'' = \frac{\omega C_{eff} R_{peff}^2}{1+\omega^2 C_{eff}^2 R_{peff}^2} \,; \qquad (1)$$

The values listed in the **Table (2)** obtained after fitting the above eqn. (1) with experimental data. Fitting parameters and detailed model are discussed in supplementary S1.

Table (2) Electrical parameters of the Au-Nanonetworks and the sample 5

| Sample | Series Resistance(KΩ) **$R_{seff}$** | Effective Capacitance(pF) **$C_{eff}$** | Effective parallel Resistance(KΩ) **$R_{peff}$** |
|---|---|---|---|
| MEG | 320.3 | 1.2 | 540.7 |
| Sample 5 | 4.4 | 13.7 | 25.5 |
| Au network | 4.3 | 12.5 | 26.7 |

Sample 5 contains contributions from the Ethylene Glycol (the base liquid) as well from the Au network. The contribution of the MEG matrix can be estimated into the model using the data of pure MEG. The value of conductivity of $3 \times 10^{-6}$S/cm obtained for MEG is very close to the standard value. The analysis of the data gives us the contribution of the Au-network which as expected makes the dominant contribution. The data presented in Table 2, show that there is a tenfold increase in the dielectric constant of the composite due to the presence of the Au nanochains. The conductivity of the nanochain is nearly 80-100 times more than that of the MEG and approaching fractions of mS/cm (Generally, in most solids when the conductivity is more than 1-10mS/cm, they show metallic behaviour). The electrical characteristics thus show clearly that there is large change due to formation of the Au nanochains.

Enhancement of the dielectric constant as well as the electrical conductivity of such a nanocomposite consisting of the Au nanoparticle chains, over that seen in MEG is a new observation. The limiting low frequency (f < 1GHz) dielectric constant $\varepsilon_0$ of MEG at room temperature is nearly 40 which in Diethylene Glycol is nearly 30 and decreasing further on polymerization [18]. The tenfold enhancement on the Au nanoparticles chain formation would mean that the nanocomposite has a dielectric constant approaching 400. This is indeed a large d.c dielectric constant. The enhancement of the dielectric constant on Au embedding in the Ethylene Glycol likely occurs due to the large polarizability the metal inclusion can provide. It has been seen that fillers with delocalized electrons can enhance the dielectric constants of polymers many fold [19, 20]. The exact cause of the enhancement of dielectric constant, however, remains a subject of future investigations.

The substantial enhancement of conductivity arising out of the Au nanoparticles inclusion at a rather low volume fraction merits attention. In this case the material consisting of Au nanoparticles interlinked by Ethylene Glycol molecules act as a network of metal electrodes connected by weakly conducting links. These molecules themselves have a low but finite conductance (G). The charge thus can hop from one metal nanoparticle to the other through these link molecules. In recent year's conductance of a number of molecules have been measured. For most molecules of the type and length used here have the conductance $G \sim 10^{-3} G_o$ where $G_0 = (2e^2/h)$. The capacitance of the Au nanoparticles connecting the molecular links (assuming them to be spheres of diameter d) is $C \approx 2\pi\varepsilon_0 d = 5x10^{-19}$ farad. The approximate local charge relaxation time in these systems $\tau_{local} = C/G \sim 5$ psec. The measured dielectric relaxation time in pure MEG, $\tau_{dielectric} \sim 100$ psec for MEG increasing to 130psec [18] for the Diethylene Glycol. The presence of the Au nanoparticles thus enhances the local charge relaxation substantially enhancing the charge hoping rate and the resulting conductivity of the composite.

*3. 4 NMR measurement and formation mechanism of the nanochains*

The main observation of the paper is that a precursor and surfactant-free one shot method can indeed produce a chain like pseudo 1D structure that spans over a large distance scale. The formation mechanism has two distinct steps, although, they can occur almost simultaneously. First is the laser ablation of the Au target that produces Au nanoparticles. Second is the arrangement of these nanoparticles into chains. The exact formation mechanism is under investigation. However, we would like to suggest some preliminary pathway that would seem likely. The formation of Au nanoparticles (diameter ~ 8-10nm) likely occurs due to ablation of the Au as droplets of similar diameters. Ablation in liquid from Au target will be similar to that seen in ablation under high pressure where there is large collision of the ablated atomic species with surrounding molecules. The heat generated during the ablation will create a vapour bubble of the surrounding liquid around the ablated region. The high surrounding pressure generally leads to the formation of Au vapour cloud and a pressure region near the surface of the Au substrate. Such larger pressure is known to produce surface hydrodynamic instabilities of the molten top Au layer from which ablation occurs and this result in the ablation of liquid in the form of droplets of similar diameters [21].

The second step , which is the formation of the chains, needs the presence of MEG and its absence, as shown above, leads to very different agglomerated geometry of the ablated Au nanoparticles. If the Au nanoparticles so formed by laser ablation arrange in chain and this is facilitated by the presence of MEG, then the nanoparticles need to first attach to the ethylene glycol. Recently it has been shown that ethylene glycol has an extraordinary capability of selectively retaining Au [22]. An investigation of an Au-PEG hybrid system has shown that the likely cause of this ethylene glycol-Au affinity has a chemical basis which arises due to strong interaction of the C-1s orbital (in the MEG) and the 4f orbitals of Au [23]. We propose that this interaction is responsible for the binding of the Au nanoparticles formed by the ablation process to ethylene glycol. This also prevents the agglomeration of the Au into larger lumps, like what happens in DI water where the ablation leads to agglomeration of the nanoparticles. The chain formation appears to be facilitated by formation of short lengths of PEG dimers or trimers by the laser irradiation. We find evidence of such a process using $^{1}$H-NMR spectroscopy. (See supplementary information S2 for details). The usual 1D-NMR spectrum (500MHz) in $D_2O$ showed the presence of proton resonance from the $CH_2$ groups of ethylene glycol. However, to check the initiation of polymerization we used 2D-NMR Correlation spectroscopy (DQF-COSY). The 2D DQF-COSY experiment was performed after drying the base liquid of the sample by lyophilizing it for one day. Cross-correlations in the spectrum allowed us to identify two terminal repeat units of PEG [-(O-$CH_2$-$CH_2$)$_n$-O-] with chemical shift dispersion indicating functionalization. Integration of the 1D-spectra with inclusion of additional peaks provided evidence for the formation of dimers and / or trimers. Since a good amount of ethylene glycol is attached to Au nanoparticles, even lyophilising the unattached MEG for 24 hours retained enough material on which NMR could be carried out. The length of MEG molecule is typically 0.27-0.28nm depending on the conformation. The average distance between two Au nanoparticles (being equal to that of a dimer/trimer) is thus ~0.55-0.8nm. This is of the same order as the persistence length of the PEG chain (~0.6nm). A schematic representation of the proposed chain is shown in figure 7. The proposed structure of one of the section of the Au networked structure is shown 7 (a) and the polymeric structure of the interlinking chain is given in figure 7 (b)

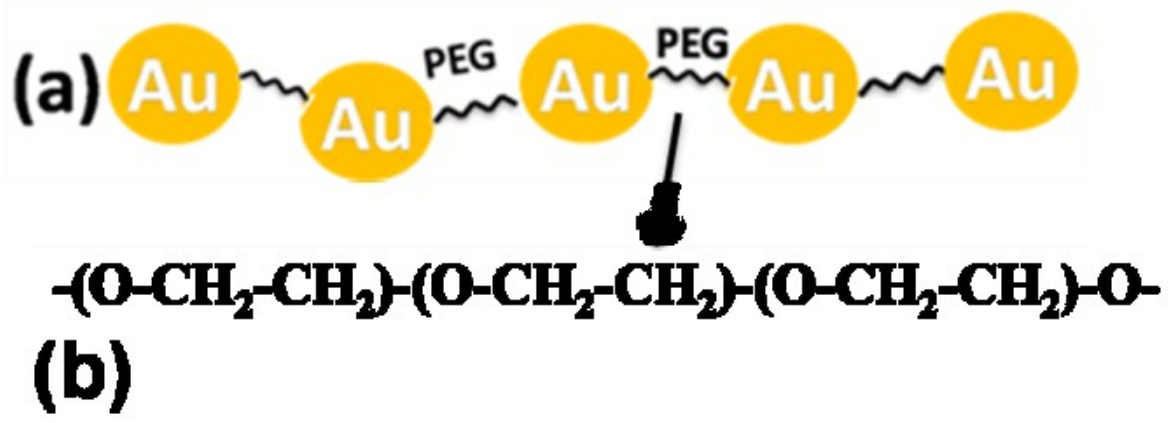

Figure 7 (a) Schematic of the proposed chain formation and (b) the polymeric structure (PEG) attached to the Au nanoparticles

The extent of polymerization during the Laser irradiation can be found out from simple analysis of the areas under the peaks of the lines belonging to the main group (MEG) at δ=3.67ppm and the peaks corresponding to the end groups. (See supplementary note 2). From this analysis it is found that the number of main MEG group is approximately 3 times that of the side groups. Thus every 4$^{th}$ molecule takes part in the polymerization and eventual chain formation.

## 4. Conclusions

We have presented a simple one step and one-pot synthesis method for making stable assembly of Au nanoparticle (diameter 8-10nm) into chains in an Ethylene Glycol medium. The method reported uses only a solid metallic Au target and a pulsed excimer laser ($\lambda$=248nm).The method does not use any precursors or reducing agent and is a chemistry free synthesis route and the complete synthesis can be done well within 15 mins without using toxic materials. We have shown that the Au nanochains can be transformed on to a solid substrate without any change. This method also provides a very simple and elegant way to convert nanoparticles to organize and form a chain like networked structure. The composite so formed shows a broad absorption in the visible region unlike sharp absorption spectrum seen in fluids with individual Au nanoparticles. Using broad frequency range impedance spectroscopy, we have successfully separated the contribution of the effective capacitance and resistance of the Au nanoparticle–Ethylene Glycol nanochains. The facilitating role of the Ethylene Glycol and its essential role in the chain formation have been established. We suggested that the quasi-1D chain like structure forms due to formation of dimer and trimers of Ethylene Glycol molecules that attach to the Au nanoparticles formed by the ablation process and facilitate the chain formation. It was found that the presence of the Au chains in the base liquid can cause an enhancement by a factor of 10, in relative dielectric constant of the sample and increases the conductivity by a factor of nearly 100 for which it may be used as capacitive materials.

## Acknowledgments

The authors acknowledge the financial support from the Department of Science and Technology and CSIR, Government of India. The authors would like to thank Dr. Samir Kr. Pal & Mr. A. Makhal for their help, coordination and fruitful discussions. Also like to thanks Dr. K Das for taking the SEM images and suggestions. IACS, Jadavpur, Kolkata for providing the TEM facility & Bose Institute, Kolkata for NMR spectroscopy.

## Supplementary Note 1(S1)

### Electrical characterisation by Impedance Spectroscopy (IS)

Impedance spectroscopy is a well established method to investigate electrical properties of materials particularly multiphase systems. Nanofluids are heterogeneous mixtures of biphasic type compost for a continuous phase (MEG) and a dispersed phase (Au network). The measurement of the small-signal ac frequency response of our AuNF sample solution and base liquid MEG has been done for wide range of frequencies, i.e., we have taken the impedance spectroscopy .Then to analysis our experimental data , we take a R-C parallel equivalent circuit model. From the equivalent circuit elements the physical parameters of the two systems are calculated. IS records the response of a system to a small applied perturbation (i.e. ac signal), over a pre-determined frequency range. The applied ac voltage and the resultant ac current are measured and the impedance calculated (Z = V/I). The technique is non-destructive and is particularly sensitive to small changes in the system. The resultant current has the same frequency f as the applied voltage but different phase θ and amplitude |Z|. By measuring the complex impedance ($Z = Z' + i Z''$) and separating the real and imaginary terms ($Z'$ and $Z''$). The complex plane plot `Nyquist plot' ($Z'$ vs. $Z''$), the capacitance dispersion C (ω) and resistance dispersion R (ω) are derived.

We have constructed the Cole-Cole plot for both the sample and the base liquid (100% MEG) which are semi circular arc in Z plane without lowering indicating a Debye relaxation process and can be modelled by equivalent electrical circuit of the type RC in parallel based on only one electroactive component in the form of semicircle. For each RC element, the complex impedance of the resistor and capacitor in parallel can be described as:

$$Z = \frac{R}{1 + j\omega CR} = \frac{R}{1 + j\omega\tau} \qquad (1)$$

Where τ=RC is the time constant of the RC element and RC element of the equivalent circuit used to represent an electroactive region within the sample will result in a semicircular arc in the complex impedance plane, Z plot. The frequency at which the semicircular arc maximum occurs is determined by the time constant t of the parallel RC element as described by: $\omega_{\max} = \tau^{-1} = (RC)^{-1}$, where $\omega_{\max}$ is the angular frequency at the top of the semicircular arc. The experimental and fitted Cole-Cole plot of base liquid MEG and AuNF sample solution is shown in Figure 7a, b (main paper).

To extract out the physical parameter like R, C etc of our sample we have to fit it with some model with the help of simulation. The model suitable for our system is shown in Fig.(9)

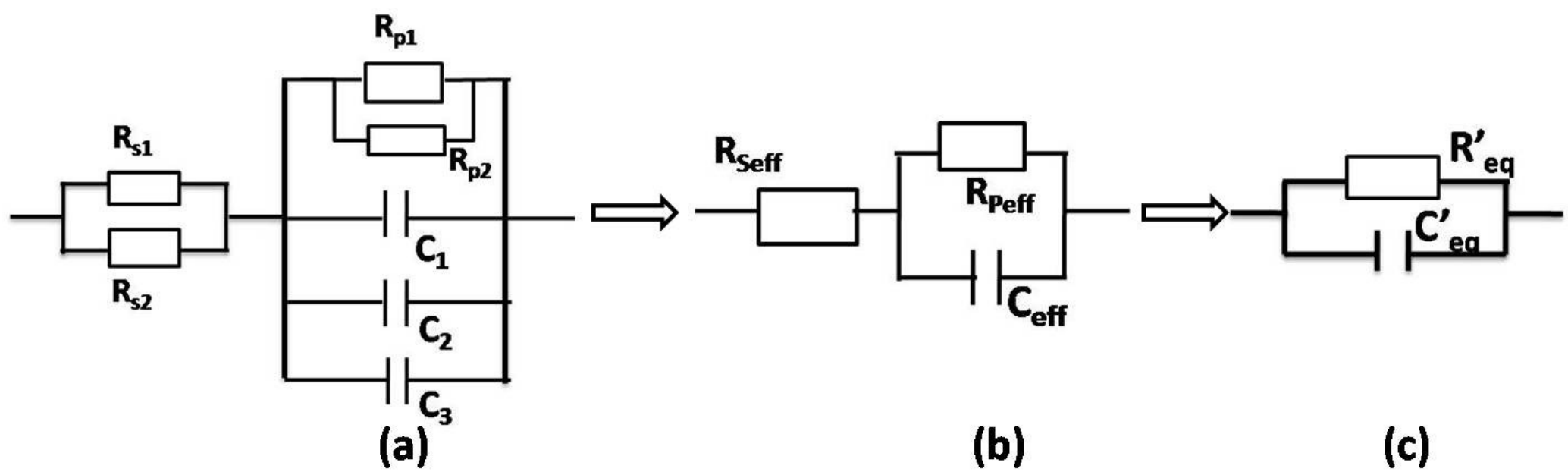

Figure S1 (a) shows the model circuit, (b)&(c) are reduced RC equivalent circuit of (a)

For MEG, $R_{seff} = R_{s1}$ = series resistance of MEG,

$R_{peff} = R_{p1}$ = parallel resistance of MEG and

$C_{eff} = C_s = (C_1+C_2)$; where $C_1$ = the capacitance of MEG.

So the real ($Z'$) and imaginary ($Z''$) part of the impedance is given below from fig.9 (b)

$$Z'_{MEG} = R_{s1} + \frac{R_{p1}}{1+\omega^2 C_s^2 R_{p1}^2}; \qquad Z''_{MEG} = \frac{\omega C_s R_{p1}^2}{1+\omega^2 C_s^2 R_{p1}^2} \tag{2}$$

From Fig.9(c) we calculate the resistive part of the Fig. 9(b) of MEG which is given below

$$R_{eq} = R_{s1} + \frac{R_{p1}}{1+\omega^2\tau^2}; \tau^2 = \frac{R_{s1}}{R_{s1}+R_{p1}} R_{p1}^2 C_s^2 = R^2 C_s^2 \tag{3}$$

Now varying $R_{s1}$, $R_{p1}$, R, $C_s$ we have fitted the experimental Cole-Cole plot of MEG and from the fitting curve $R_{s1}$, $R_{p1}$ & $C_s$ of MEG (base liquid) are obtained and listed in Table 1.

For sample 5, $R_{seff} = R_{s1} \| R_{s2}$; where $R_{s2}$ = series resistance of the Au nanonetworks

$R_{peff} = R_{p1} \| R_{p2}$; where $R_{p2}$ = parallel resistance of the Au nanonetworks and

$C_{eff} = C_1 + C_2 + C_3$; where $C_3$ = capacitance of the Au nanonetworks.

So from Fig. 9(b), we find out the real ($Z'$) and imaginary ($Z''$) part of the sample solution and they are given below

$$Z'_{sample} = R_{seff} + \frac{R_{peff}}{1+\omega^2 C_{eff}^2 R_{peff}^2} \quad ; Z''_{sample} = \frac{\omega C_{eff} R_{peff}^2}{1+\omega^2 C_{eff}^2 R_{peff}^2} \; ; \tag{4}$$

From the Fig. 9(c) we calculate the resistive part of the Fig. 9(b) of the sample and it is given below

$$R'_{eq} = R_{seff} + \frac{R_{peff}}{1+\omega^2 \tau'^2} ; \tau'^2 = \frac{R_{seff}}{R_{seff}+R_{peff}} R_{peff}^2 C_{eff}^2 = R'^2 C_{eff}^2 \tag{5}$$

Now varying the $R_{seff}$, $R_{peff}$, $C_{eff}$ we have fitted the experimental Cole-Cole plot of Sample solution and from the fitting curve we find out the $R_{seff}$, $R_{peff}$ and $C_{eff}$ and from these quantities we can easily find out $R_{s2}$, $R_{p2}$, $C_3$(in the main paper which are denoted as $R_{seff}$, $R_{peff}$, $C_{eff}$ ) of the Au network Table(2) (in main paper) .So from the fitting results of our Aunf and base liquid MEG samples we can extract out the dielectric constant, effective capacitance and resistances of the two phases of the composite separately. Values of the parameters of the samples coming out after fitting are noted in the Table (2).

## Supplementary Note 2 (S2)

### NMR experiment

$^{1}H$-NMR experiments were performed in $H_2O/D_2O$ with TSP as the internal standard on a Bruker DRX 500 spectrometer at 25°C. $^{1}H$-NMR spectrum of an aliquot of MEG (not shown), after laser ablation of Au, showed a single peak at 3.67 ppm (corresponding to pure MEG) indicating that the major solvent component was still MEG. However, when the solvent aliquot was lyophilized (24 hours) to remove MEG and the residue re-dissolved in $D_2O$, in addition to the original signal (singlet) at 3.67 ppm, additional signals were observed (shown in Fig. S1). In addition to a complex multiplet (3.7-3.8 ppm), four triplets were observed at 2.9, 3.65, 3.89 and 4.78 (the latter overlapped with residual water signal) ppm. 2D DQF-COSY experiments showed the four triplets to be connected as two pairs: 2.9-3.89 and 3.65-4.78. The appearance of two pairs of J-coupled triplets is consistent with what is expected from two di-methylene units in the basic repeat unit in PEG [-(O-$CH_2$-$CH_2$)$_n$-O-]. The chemical shift dispersion among the methylene protons in the two di-methylene units further indicated that the two units correspond to the termini of PEG, linked to the Au nanoparticles. The complex multiplet (3.7-3.8 ppm) either originates from the central repeat unit of a PEG trimer (integral of the multiplet close to double the integral of one triplet), or represents other variants of PEG. In summary, NMR data show: i) formation of PEG upon laser ablation, ii) functionalization of PEG (chemical shift dispersion between J-coupled triplets), iii) formation of dimers (the two pairs of J-coupled triplets) or the formation of trimers (if the multiplet is due to another repeating unit (sandwiched between the terminal repeats).

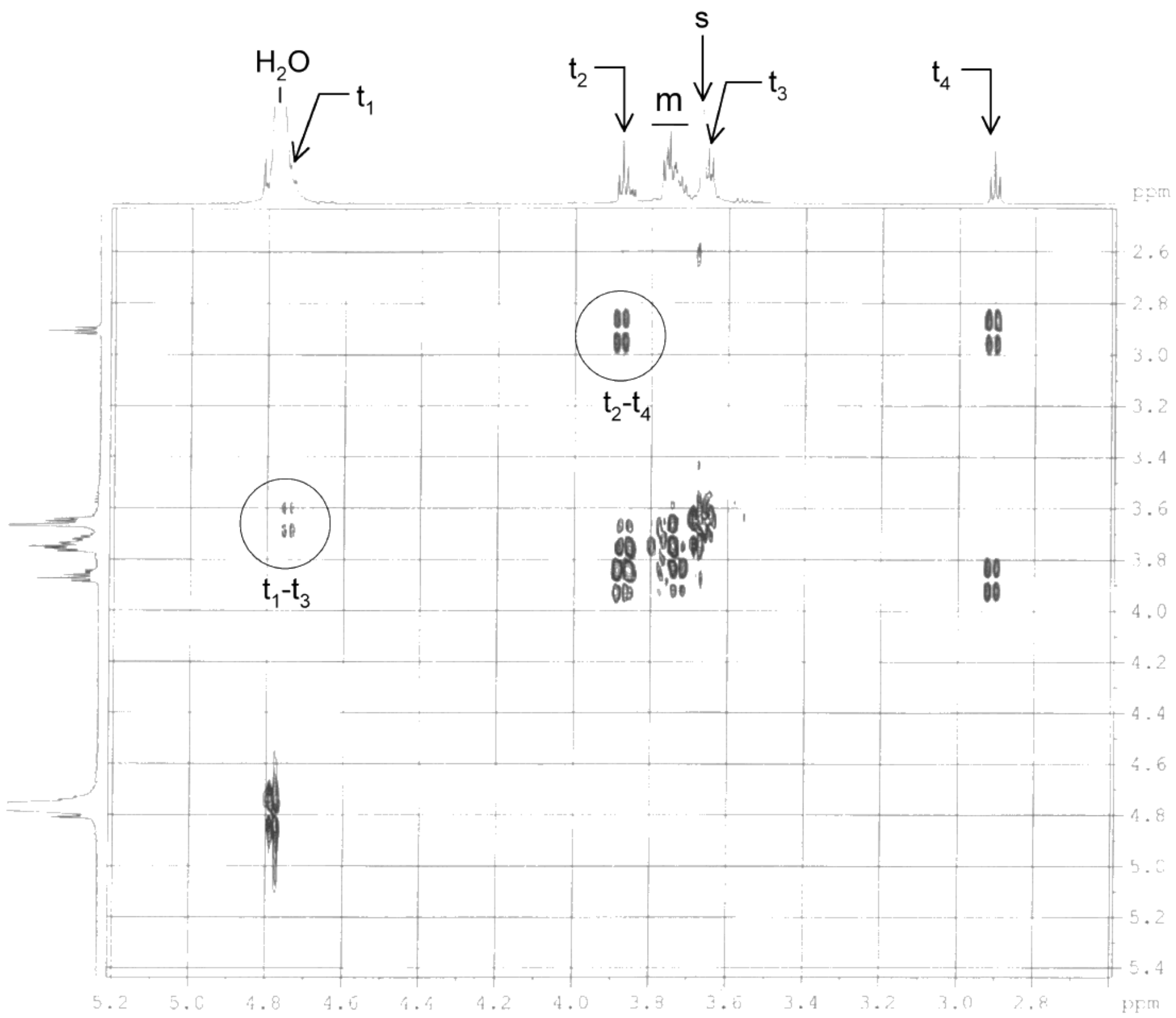

**Figure S2.1** DQF-COSY spectrum of lyophilized solvent (MEG) aliquot after laser ablation of Au in $D_2O$ at 25°C. The 1D spectra are shown along the two axes. Cross peaks between the triplets, marked as t1, t2, t3 and t4, are circled and annotated. The singlet (s) was assigned to residual MEG and the multiplet (m) was assigned to the central repeat unit of MEG trimer.